\documentclass[a4paper]{article}
\usepackage{epsfig,amssymb,amsmath}

\def\beq{\begin{eqnarray}}
\def\eeq{\end{eqnarray}}
\def\bsp{\begin{split}}
\def\esp{\end{split}}

\newcommand{\mf}[1]{{\mathfrak #1}}

\newcommand{\mb}[1]{{\mathbb #1}}
\newcommand{\mc}[1]{{\cal #1}}
\newcommand{\mbold}[1]{\mbox{\boldmath{\ensuremath{#1}}}}

\begin{document}

\title{\textbf{Multidimensional Cosmology: \\ Spatially Homogeneous
models of dimension 4+1}}
\author{\textbf{Sigbj\o rn Hervik}\thanks{S.Hervik@damtp.cam.ac.uk}
\\ \\
 DAMTP,\\
Centre for Mathematical Sciences,\\
Cambridge University, \\
 Wilberforce Rd.,\\
Cambridge CB3 0WA, UK}
\date{\today}
\maketitle

\begin{abstract}
In this paper we classify all 4+1 cosmological models where the
spatial hypersurfaces are connected and  simply connected homogeneous
Riemannian manifolds. These models come in two categories, multiply transitive and
simply transitive models. There are in all five different multiply
transitive models which cannot be considered as a special case of a
simply transitive model. 
The classification of simply transitive models, relies heavily upon
the classification of the four dimensional (real) Lie algebras. For
the 
orthogonal case, we derive all the equations of motion and give some examples of exact solutions. 
Also the problem of how these models can be compactified in context with the Kaluza-Klein
mechanism, is addressed.
\end{abstract}

\section{Introduction}
The idea that our world has more dimensions than three is
actually  older than the theory of general relativity itself. Already
in 1914, G. Nordstr\"om\footnote{English
translations of this and  other related works  are
found in the book \cite{KaluzaKlein}.}  put forward a five-dimensional
scalar-tensor theory in an effort to unify gravity and
electromagnetism. Since it was based upon his own theory of
gravitation which was soon superseded by Einstein's theory, this work
was neglected for several decades. 

However, in 1919, T. Kaluza \cite{kaluza}  constructed a similar unified theory
of gravity and electromagnetism based on the linearized version of the
general theory of relativity. 
 Kaluza's work, which was published in 1921 and  was followed by two important papers by  Klein
\cite{klein1,klein2}, had a very interesting result: five-dimensional Einstein
gravity could be seen upon as
Einstein gravity in four dimensions plus electromagnetism. 

The  original idea had only one extra dimension, but string
theorists today believe that we can have up to seven extra
dimensions\cite{polchinski}. Supposedly,  six of these have to be curled up in a small
Calabi-Yau manifold\cite{green}. Hence, among theorists the question is not
\emph{if} we have extra dimensions, but rather ``\emph{How many extra
dimensions are there?}'' and ``\emph{What is their nature?}''

In this paper we will consider the simplest models, models with one
extra dimension. We will investigate cosmological models of 4+1
dimensions. In 3+1 dimensions the classification relies on Bianchi's
classification of the homogeneous three-manifolds
\cite{bianchi}. Bianchi's work was later generalised to the
four-dimensional manifolds by Fubini \cite{Fubini}. In
three dimensions the homogeneous manifolds have a special
role. According to the Thurston conjecture, the homogeneous manifolds
of dimension three are intimately related to the classification of
three-manifolds\cite{thurston97,thurston}. There does not exist a similar conjecture in four
dimensions; on the contrary, topology in four dimensions is
completely different. Notwithstanding, we will
assume that our 4+1 cosmological model is \emph{spatially
homogeneous}. This assumption heavily restricts the number of possible four-manifolds
to only a finite and manageable number. We know
that our universe is homogeneous on scales larger than a billion
light-years, and thus the assumption of homogeneity is by no means a
radical and unrealistic one\footnote{From CMB measurements we know
that the fluctuations away from homogeneity are no larger than
$10^{-5}$ at the last scattering surface.}. 

The purpose of our study is to investigate the impact of extra
dimensions on the cosmic evolution of our universe. This may give us
some understanding of how extra dimensions generically influence the
 observed 3+1 dimensional universe. 

The simplest extension we can think of, going from three to four
spatial dimensions, is just to assume that we have have a product
space:
\beq
\Sigma\cong M\times S^1
\eeq
where $\Sigma$ is a four dimensional spatial manifold and $M$ a three dimensional manifold. However, this naive
assumption does not necessarily need to be true. Experimental data from
particle accelerators indicate
 that the size of this small extra dimension must be less than
$10^{-18}$m. Hence, since the size of the base-space $M$ is greater than billions of light-years, the extra dimension may also
be ``twisted''. This twisting is at a global scale, and will therefore
be unmeasurable in particle accelerators. However, we know
that once in the past, the size of the universe might have been
comparable with the small extra dimensions. Actually, it is one of
our principal aims to try to explain why one (or more) dimension(s) is so
incredibly much smaller than the three large ones we see today. 

Some work along similar lines already exist in the literature. Forgacs
and Horvath \cite{FH1,FH2} investigated already in 1979 how higher
dimensional models could  influence
the cosmology of isotropic
FRW universes. This work was followed by Chodos and
Detweiler\cite{CD} who studied the translational invariant
higher-dimensional cosmological models -- also over 20 years ago. Later, other
people have investigated other models
\cite{Ish,johnd,DHHS,DHS,DSS,ADIP,HK,McM,CMcM,ACM1,ACM2,SH,halpern2,halpern3,CGHW}.
Also, Lorenz-Petzold produced a string of papers
\cite{LP1,LP2,LP3,LP4,LP5,LP6,LP7,LP8} by lifting the
homogeneous models in 3+1 dimensions, one by one, to higher
dimensions. Maybe the closest work related to this
(at least to the authors knowledge) are two articles in the
mid-eighties \cite{DH,halpern}. However, there does not seem to be any
work which tries to classify all the spatially homogeneous spacetimes in 4+1
dimensions and gives a systematic approach to the equations of
motion. The aim of this paper is exactly to do this; give a
classification and with the aid of the orthonormal frame formalism, we
will derive all the equations of motion. 

The paper is organised as follows. First we classify all the multiply
transitive spaces. We will throughout our paper assume, unless
stated otherwise, that our space is connected and simply connected. It
is of special interest to find all the homogeneous spaces
which cannot be considered as simply transitive spaces. The simply
transitive spaces are classified in  section \ref{sect:simply}. This
classification heavily relies upon the classification of the
four-dimensional Lie algebras. We write down all the equations of
motion and go on and provide with some examples and give some exact
solutions. Lastly, we go to the question of compactification
which  is one of the key ingredients of the Kaluza-Klein
mechanism. 

\section{Multiply transitive models}
Let us first classify all the multiply transitive models which are
connected and simply connected. In dimension
three there is only one multiply transitive model which cannot be
considered as a special case of a simply transitive one. This is the
well-known Kantowski-Sachs model. The symmetry group in this case is
four-dimensional and it has  three-dimensional subgroup which acts on a
two-sphere $S^2$. 

Our analysis is based on the classification of the homogeneous Riemannian
spaces of dimension four due to Ishihara \cite{ishihara}\footnote{The homogeneous
Riemannian spaces are treated in a more general way in for example \cite{kob}. However, dimension
four seems to be a special case. This is mainly due to the non-simpleness
of $SO(4)$.}.  Large parts of our results in this section can be
extracted directly from this paper. In four dimensions there are in all 5
different models which cannot be seen as special cases of a
simply transitive space. In the following, these will be
emphasized.\footnote{The underlined spaces will correspond to these
multiply transitive models. However, they will only be underlined in
the section were they have the smallest transitive symmetry group.}

\subsection{Maximally symmetric: dim\,Isom$(\Sigma)=10$}
There are three (orientable) maximally symmetric spaces of dimension
4. They are the well-known cases: $\underline{S^4},~\mb{E}^4$ and
$\mb{H}^4$. These have an isotropy group isomorphic to $SO(4)$. They
correspond to the three different FRW cosmological models in 5D. Both
the Euclidean space and the hyperbolic space have subgroups of the
isometry group which acts simply transitive on their respective
spaces. $\underline{S^4}$ on the other hand, has no proper subgroup that acts
transitively on the space at all. Hence, $\underline{S^4}$ cannot be considered as
a special case of a simply transitive space. 

\subsection{K\"ahler manifolds: dim\,Isom$(\Sigma)=8$} 
Interestingly, there also exist four-dimensional Riemannian spaces
which have an 8-dimensional isometry group\footnote{In general for a
homogeneous Riemannian space $M$ of dimension $n\neq 4$ there does not exist
a closed subgroup of Isom$(M)$ of dimension $r$ such that
$n(n+1)/2>r>1+n(n-1)/2$. These are the counterexamples for $n=4$.}. These are the three
K\"ahler manifolds of constant curvature:
$\underline{\mb{C}\mb{P}^2},~\mb{C}^2$ and\footnote{For those who are unfamiliar
with the complex hyperbolic spaces, consult for example Goldman's book
\cite{goldman}.} ${\mb{H}_{\mb{C}}^2}$. These spaces have an isotropy
subgroup isomorphic to $U(2)$. Both  $\mb{C}^2$ and ${\mb{H}_{\mb{C}}^2}$
have a simply transitive subgroup. 

The Riemann curvature tensor for these spaces can be written in
complex coordinates as
\beq
R_{\alpha\bar{\beta}\delta\bar{\gamma}}=K\left(g_{\alpha\bar{\beta}}g_{\delta\bar{\gamma}}+g_{\alpha\bar{\gamma}}g_{\delta\bar{\beta}}\right)
\eeq
where $K>0$, $K=0$ and $K<0$ for ${\mb{C}\mb{P}^2},~\mb{C}^2$ and
${\mb{H}_{\mb{C}}^2}$ respectively. Thus these spaces have constant
holomorphic sectional curvature. The isotropy group is the group of
motions that leaves the K\"ahlerian structure invariant. 

\subsection{dim\,Isom$(\Sigma)=7$}
Spaces that possess a 7-dimensional isometry group come in three
classes. The first class consists of maximally symmetric three-manifolds times a
line: $\mb{E}^1\times S^3, \, \mb{E}^4$ and
$\mb{E}^1\times\mb{H}^3$. These have an isotropy group isomorphic to
$SO(3)$. The second category consists of $\mb{H}^4$ with a certain
7-dimensional symmetry group. The third is the K\"ahler manifold 
$\mb{C}^2$ with isotropy group $SU(2)$. All of these spaces can be
considered as a special case of a simply transitive space. 

\subsection{dim\,Isom$(\Sigma)=6$}
All of these spaces are products of two two-dimensional maximally
symmetric spaces: $\underline{S^2\times S^2},\, S^2\times\mb{E}^2,\,
S^2\times\mb{H}^2,\, \mb{H}^2\times\mb{E}^2,\,\mb{H}^2\times\mb{H}^2$
and $\mb{E}^4$. These have the isotropy group $SO(2)\times
SO(2)$. Only $\underline{S^2\times S^2}$
does not have a lower dimensional subgroup of the isometry group which
acts transitively on the spatial hypersurface. For this model, the
metric can be written as
\beq
ds^2=-dt^2+a(t)^2\left(d\theta^2+\sin^2\theta d\phi^2\right)+ 
b(t)^2\left(d\psi^2+\sin^2\psi d\xi^2\right).
\eeq

\subsection{dim\,Isom$(\Sigma)=5$}
There are several spaces possessing a 5-dimensional isometry
group. Only two of them cannot be considered as a special case of a
simply transitive space: $\underline{S^2\times\mb{E}^2}$ and $
\underline{S^2\times\mb{H}^2}$. The isotropy group for both these spaces is
$SO(2)$. Examples of such models for the case $\underline{S^2\times \mb{E}^2}$
case are
\beq
ds^2=-dt^2+a(t)^2\left(d\theta^2+\sin^2\theta d\phi^2\right)+ 
b(t)^2\left(e^{-2\beta(t)}dx^2+e^{2\beta(t)}dy^2\right),
\eeq
and in the case  $\underline{S^2\times \mb{H}^2}$
\beq
ds^2=-dt^2+a(t)^2\left(d\theta^2+\sin^2\theta d\phi^2\right)+ 
b(t)^2\left(e^{-2\beta(t)}dx^2+e^{2\beta(t)+2Kx}dy^2\right).
\eeq
The functions $a(t)$, $b(t)$ and $\beta(t)$ will be
determined by the field equations. 
 
\section{Simply transitive models}
\label{sect:simply}
In the simply transitive case, we can systematically construct the
homogeneous spaces using the classification of the 4-dimensional Lie
algebras. For a given simply transitive space, the Killing vectors
obey a certain commutator relation
\beq
[{\mbold\xi}_i,{\mbold\xi}_j]=\widetilde{C}^k_{~ij}{\mbold\xi}_k
\label{Ctilde}\eeq
where the structure constants $\widetilde{C}^k_{~ij}$ are functions of
$t$ only. In the following we will follow Ellis and MacCallum \cite{EM}
closely. However we will use a slightly different notation: Greek
indices ($\mu,\nu,...$) have range 0-4 over the full five dimensional
space-time; lower case Latin indices ($a,b,i,j,...$) have range 1-4
over the four-dimensional spatial hypersurfaces; upper case Latin
indices ($A,B,...$) have range 1-3 over three spatially directions. We
will assume that the model is orthogonal, i.e. we assume that our
spacetime is of the form
\beq
\Sigma_t\times \mb{R}
\eeq
where the fluid four-velocity ${\bf u}$ is orthogonal to $\Sigma_t$. 

We can now introduce a left-invariant spatial frame ${\bf e}_a$, which together
with the fluid four-velocity ${\bf u}={\bf e}_t$ forms a orthonormal frame denoted by
${\bf e}_{\mu}$. These commute, by definition, with the Killing vectors:
\beq
[{\mbold\xi}_i,{\bf e}_{\mu}]=0.
\eeq
The f\"unfbein ${\bf e}_{\mu}$ will now obey the commutation relations
\beq
[{\bf e}_{\mu},{\bf e}_{\nu}]=\gamma^{\rho}_{~\mu\nu}{\bf e}_{\rho}.
\eeq
We can relate these commutator function to the connection coefficients
for the particular orthonormal basis. For an orthonormal basis we
define the rotation forms by
\beq
{\bf de}_{\mu}={\bf e}_{\nu}\otimes{\mbold\Omega}_{~\mu}^{\nu}.
\eeq
Denoting the dual frame of ${\bf e}_{\mu}$ by ${\mbold\omega}^{\mu}$,
the rotation forms can be related to the connection coefficients via
\beq
{\mbold\Omega}_{~\mu}^{\nu}=\Gamma^{\nu}_{~\mu\lambda}{\mbold\omega}^{\lambda}.
\eeq
Also, they possess the antisymmetry ${\mbold\Omega}_{\mu\nu}=-{\mbold\Omega}_{\nu\mu}$.
Hence, the connection coefficients $\Gamma^{\rho}_{~\mu\nu}$ for
the orthonormal frame possess the antisymmetry
\beq
\Gamma_{\rho\mu\nu}=-\Gamma_{\mu\rho\nu}
\eeq
and can be written as
\beq
\gamma^{\rho}_{~\mu\nu}&= &-\left(\Gamma^{\rho}_{~\mu\nu}-\Gamma^{\rho}_{~\nu\mu}\right)\nonumber
\\
\Gamma_{\rho\mu\nu}&=& -\frac
12\left(\gamma_{\rho\mu\nu}+\gamma_{\mu\nu\rho}-\gamma_{\nu\rho\mu}\right).
\eeq

The orthogonality condition requires that
\beq
\gamma^0_{~0a}=\gamma^0_{~ab}=0.
\eeq
We can split the remaining part of the structure constants into 
\beq
\gamma^a_{~0b}&=&-\theta^a_{~b}+\Omega^a_{~b}\nonumber \\
\gamma^i_{~jk}&=&C^i_{~jk}.
\eeq
Here are $\Omega_{ab}$ the angular velocity in the $ab$-plane of a
Fermi-propagated axis with respect to the triad ${\bf e}_a$, and
$\theta^a_{~b}$ the volume expansion tensor. If $u^{\mu}$ is the
time-like vector-field orthogonal to the hypersurfaces $\Sigma_t$, then
$\theta^a_{~b}$ is defined by $\theta_{\mu\nu}=u_{\mu;\nu}$. One can
readily see that  $\theta_{\mu\nu}$ is symmetric and
$\theta_{\mu\nu}u^{\mu}=0$. 

We can further split the volume expansion tensor into a trace and
trace-free part
\beq
\theta_{\mu\nu}=\frac{1}{4}h_{\mu\nu}\theta+\sigma_{\mu\nu}.
\eeq
The tensor $\sigma_{\mu\nu}$ is the shear tensor and
$h_{\mu\nu}=g_{\mu\nu}-u_{\mu}u_{\nu}$ is the projection tensor
onto the hypersurfaces $\Sigma_t$. 

 The
structure constants $C^k_{~ij}$ defines the specific Lie algebra under
consideration. 
These can be separated into a trace (or vector) part
and a trace-free part
\beq
C^k_{~ij}=D^k_{~ij}+\delta^k_{~i}a_j-\delta^k_{~j}a_i
\eeq
where
\beq
a_i&=& \frac 13{\rm Tr}(C^k_{~ji})\equiv \frac 13 C^j_{~ji}\nonumber
\\
D^k_{~ij} &=& C^k_{~ij}-\frac 23 C^l_{~[i|l|}\delta^k_{~j]}.
\eeq
The Lie algebra given by the structure constants $C^k_{~ij}$ is
isomorphic to the Lie algebra $\widetilde{C}^k_{~ij}$ defined in eq. (\ref{Ctilde}).

The Jacobi identity,
\beq
[{\bf e}_{\mu},[{\bf e}_{\nu},{\bf e}_{\rho}]]+[{\bf e}_{\nu},[{\bf
e}_{\rho},{\bf e}_{\mu}]]+[{\bf e}_{\rho},[{\bf e}_{\mu},{\bf
e}_{\nu}]]=0
\eeq
reduces now to the following constraint equations
\beq
a_kD^k_{~ij}&=& 0\label{Jacobi1} \\
D^d_{~ab}D^e_{~cd}+D^d_{~ca}D^e_{~bd}+D^d_{~bc}D^e_{~ad}+D^e_{~ab}a_c+D^e_{~ca}a_b+D^e_{~bc}a_a&=&
0\label{Jacobi2}
\eeq
and evolution equations
\beq 
\dot{a}_b+\frac 14\theta
a_b-\left(\Omega^d_{~b}-\sigma^d_{~b}\right)a_d&=&0\label{Jacobi3} \\
\dot{D}^a_{~bc}+\frac 14
\theta{D}^a_{~bc}+2\left(\Omega^d_{~[b}-\sigma^d_{~[b}\right){D}^a_{~c]d}+\left(\Omega^a_{~d}-\sigma^a_{~d}\right)D^d_{~bc}&=&0\label{Jacobi4}.
\eeq
Equation (\ref{Jacobi2}) is just the Jacobi identity for
$C^k_{~ij}$ in disguise, while eq. (\ref{Jacobi1}) is the trace of
eq. (\ref{Jacobi2}).

\subsection{The Field Equations}
Let us assume that the energy-momentum tensor is of the form
\beq
T_{\mu\nu}=(\rho+p)u_{\mu}u_{\nu}+pg_{\mu\nu}+\pi_{\mu\nu}
\eeq
where $\pi_{\mu\nu}$ is the anisotropic stress tensor. This tensor is
symmetric and has
the properties
\beq
\pi^{\mu}_{~\mu}=u^{\mu}\pi_{\mu\nu}=0.
\eeq
The identity $T^{\mu}_{~\nu;\mu}=0$ leads to the energy conservation
equation
\beq
\dot{\rho}+(\rho+p)\theta+\pi^{\mu\nu}\sigma_{\mu\nu}=0.
\eeq
Together with an equation of state for the fluid, these equations
govern the evolution of the fluid in our model.

The Riemann curvature tensor is given by
\beq
R^{\alpha}_{~\beta\gamma\delta}={\bf
e}_{\gamma}\left(\Gamma^{\alpha}_{~\beta\delta}\right)-{\bf
e}_{\delta}\left(\Gamma^{\alpha}_{~\beta\gamma}\right)+\Gamma^{\alpha}_{~\lambda\gamma}\Gamma^{\lambda}_{~\beta\delta}-\Gamma^{\alpha}_{~\lambda\delta}\Gamma^{\lambda}_{~\beta\gamma}-\Gamma^{\alpha}_{~\beta\lambda}\gamma^{\lambda}_{~\gamma\delta}.
\eeq
From this expression we can readily calculate the field equations. The
5D Raychaudhuri's equation is
\beq
\dot\theta+\frac 14 \theta ^2+\sigma^{ab}\sigma_{ab}+\frac
23\left(\rho+2p-\Lambda\right)=0
\eeq
while the $(0,a)$ equations are
\beq
4\sigma_{ac}a^c-\sigma^b_{~c}{D}^{c}_{~ba}=0.
\label{R0a}\eeq

The trace of the field equations yields the 5D Friedmann equation (or the
constraint equation):
\beq
\frac 38 \theta ^2 &=& \frac 12 \sigma^{ab}\sigma_{ab}+6a^2+\frac
18\mf{D}^2+\rho+\Lambda \\
\mf{D}^2&=&
D^{a}_{\phantom{a}bc}D^{\phantom{a}bc}_{a}+2D^{a}_{\phantom{a}bc}D^{b\phantom{a}c}_{\phantom{b}a}\nonumber.
\eeq
The shear equations can be derived from the trace-free part of the
field equations:
\beq
\dot{\sigma}_{ab}+\theta\sigma_{ab}+\frac
32\left(D_{abd}+D_{bad}\right)a^d+2\sigma^c_{~(a}\Omega_{b)c}+b_{ab}-\frac
14 h_{ab}b^c_{~c}=\pi_{ab} 
\eeq
where we have set
\beq
b_{ab}\equiv - \frac 14\left(
2D^c_{\phantom{c}ad}D^{\phantom{cb}d}_{cb}-D_{acd}D^{\phantom{b}cd}_{b}+2D^c_{\phantom{c}ad}D^{d}_{\phantom{d}bc}\right).
\eeq
Note that $4b^{a}_{~a}=-\mf{D}^2$. The tensor $b_{ab}$ can be
interpreted as part of the spatial Ricci tensor:
\beq
{}^{(4)}R_{ab}=b_{ab}+\frac
32\left(D_{abd}+D_{bad}\right)a^d-3h_{ab}a^2.
\eeq
Hence, the curvature of the spatial four-surfaces is
\beq 
{}^{(4)}R=b^a_{~a}-12a^2=-\frac{1}{4}\mf{D}^2-12a^2.
\eeq

\subsection{The four-dimensional Lie algebras} 
In a series of papers Mubarakzyanov \cite{mub1,mub2,mub3} gave a
classification of real Lie algebras up to dimension 5.\footnote{The
classification of the \emph{complex} Lie algebras up to dimension 6 was
actually done by Sophus Lie himself already in the 1890's \cite{Lie}.} A list of all the
algebras is given in \cite{PSW,PW} which is more accessible for
non-Russian readers (see also \cite{Fee} for the four-dimensional case). Here we will be interested in the four-dimensional
classification and we will be using mostly the notation in
\cite{PSW}. A useful reference is also a report  by MacCallum
\cite{MacC}. 

We will investigate Lie algebras from two different point of views; one
from a geometric point of view, and the other from an
algebraic point of view. The two different ways of investigating Lie algebras
both have their strengths and weaknesses and are useful for different
purposes. We will start out from the geometrical point of view;
dividing them into decomposable and indecomposable ones. The
geometrical picture is completely
different in these two different classes. The decomposable ones naturally give rise to product
spaces\footnote{As Lie groups.}, while
the indecomposable ones cannot be written as a topological product.  

\subsubsection{Decomposable Lie algebras}
Since the three-dimensional Lie algebras are used quite frequently in
the literature we will use a notation similar to the Bianchi types
when the algebras are composed of these. The decomposable
four-dimensional Lie algebras are 
\beq
{{A_3}}\oplus\mb{R}, \quad\text{and}\quad A_{2,1}\oplus A_{2,1}
\eeq
where ${A_3}$ is one of the Bianchi type Lie algebras, labelled
I-IX, and $A_{2,1}$ is the only non-trivial two-dimensional Lie
algebra. In this notation, the extension of the type IX Bianchi type
will be denoted IX$\oplus\mb{R}$. $A_{2,1}$ can be represented by the single non-trivial
commutator
\beq
[{\bf e}_1,{\bf e}_2]={\bf e}_1.
\eeq
One can show that this Lie algebra acts simply transitively on $\mb{H}^2$. 
It has been known for a while that the commutators for the Bianchi Lie
algebras can be written
\beq
C^A_{\phantom{A}BC}=\epsilon_{BCD}n^{DA}+\delta^A_{\phantom{A}B}a_C-\delta^A_{\phantom{A}C}a_B
\eeq
where $n^{AD}$ is a symmetric matrix. This is called the \emph{Behr
decomposition} \cite{behr}. The Jacobi identity reduces to
the single relation
\beq
n^{AD}a_D=0.
\eeq
The Bianchi types can now be classified in terms of the eigenvalues of
the matrix $n^{AD}$ \cite{EM}.

\subsubsection{Indecomposable Lie algebras}
The classification of the indecomposable four-dimensional Lie
algebras are listed in table \ref{table4dim}.
\begin{table}
\begin{tabular}{|c|c|c|}
\hline \hline
Name & Commutator relations & $a_i$ \\
\hline \hline
$A_{4,1}$ & $[{\bf e}_2,{\bf e}_4]={\bf e}_1,\quad [{\bf e}_3,{\bf
e}_4]={\bf e}_2 $ & $0$ \\ \hline 

$A_{4,2}^p$ & $\begin{matrix} [{\bf e}_1,{\bf e}_4]=p{\bf e}_1,\quad [{\bf e}_2,{\bf
e}_4]={\bf e}_2, \\ {} [{\bf e}_3,{\bf
e}_4]={\bf e}_2+{\bf e}_3 \ \ (p\neq 0)  \end{matrix}$ & $\frac 13(p+2)\delta_{~i}^4$ \\ \hline

$A_{4,3}$ & $[{\bf e}_1,{\bf e}_4]={\bf e}_1,\quad [{\bf e}_3,{\bf
e}_4]={\bf e}_2  $ & $\frac 13\delta_{~i}^4$ \\ \hline

$A_{4,4}$ & $\begin{matrix}[{\bf e}_1,{\bf e}_4]={\bf e}_1,\quad [{\bf e}_2,{\bf
e}_4]={\bf e}_1+{\bf e}_2, \\ {} [{\bf e}_3,{\bf
e}_4]={\bf e}_2+{\bf e}_3  \end{matrix} $ & $\delta^4_{~i} $\\ \hline

$A_{4,5}^{pq}$ & $\begin{matrix} [{\bf e}_1,{\bf e}_4]={\bf e}_1,\quad [{\bf e}_2,{\bf
e}_4]=p{\bf e}_2, \quad [{\bf e}_3,{\bf
e}_4]=q{\bf e}_3 \\ (pq\neq 0,\quad -1\leq q\leq p \leq 1)
\end{matrix}$  & $\frac 13(1+p+q)\delta^4_{~i}$ \\ \hline

$A_{4,6}^{pq}$ & $\begin{matrix} [{\bf e}_1,{\bf e}_4]=p{\bf e}_1,\quad [{\bf e}_2,{\bf
e}_4]=q{\bf e}_2-{\bf e}_3, \\ {} [{\bf e}_3,{\bf
e}_4]={\bf e}_2+q{\bf e}_3 \quad (p\neq 0,\quad q \geq 0)
\end{matrix}$  & $\frac 13(p+2q)\delta^4_{~i}$ \\ \hline

$A_{4,7}$ & $\begin{matrix} [{\bf e}_2,{\bf e}_3]={\bf e}_1,\quad [{\bf e}_1,{\bf
e}_4]=2{\bf e}_1, \quad [{\bf e}_2,{\bf
e}_4]={\bf e}_2,\\ {} [{\bf e}_3,{\bf e}_4]={\bf e}_2+{\bf e}_3  \end{matrix} $ &
$\frac 43\delta^4_{~i}$
\\ \hline

$A_{4,8}$ & $[{\bf e}_2,{\bf e}_3]={\bf e}_1,\quad [{\bf e}_2,{\bf
e}_4]={\bf e}_2, \quad [{\bf e}_3,{\bf
e}_4]=-{\bf e}_3   $ & $0$ \\ \hline

$A_{4,9}^{q}$ & $\begin{matrix} [{\bf e}_2,{\bf e}_3]={\bf e}_1,\quad [{\bf e}_1,{\bf
e}_4]=(1+q){\bf e}_1, \quad [{\bf e}_2,{\bf
e}_4]={\bf e}_2,\\ {} [{\bf e}_3,{\bf e}_4]=q{\bf e}_3 \quad ( -1< q
\leq 1) \end{matrix}$  & $\frac 23(1+q)\delta^4_{~i}$ \\ \hline 

$A_{4,10}$ & $[{\bf e}_2,{\bf e}_3]={\bf e}_1,\quad [{\bf e}_2,{\bf
e}_4]=-{\bf e}_3, \quad [{\bf e}_3,{\bf
e}_4]={\bf e}_2 $ & $0$ \\ \hline

$A_{4,11}^{q}$ & $\begin{matrix} [{\bf e}_2,{\bf e}_3]={\bf e}_1,\quad [{\bf e}_1,{\bf
e}_4]=2q{\bf e}_1, \quad [{\bf e}_2,{\bf
e}_4]=q{\bf e}_2-{\bf e}_3, \\ {} 
[{\bf e}_3,{\bf e}_4 ] = {\bf e}_2+q{\bf
e}_3,  \quad ( q>0) \end{matrix}$  & $\frac 43q\delta^4_{~i}$ \\ \hline 

$A_{4,12}$ & $\begin{matrix} [{\bf e}_1,{\bf e}_4]={\bf e}_1,\quad [{\bf e}_2,{\bf
e}_4]={\bf e}_2, \\ {} [{\bf e}_1,{\bf
e}_3]=-{\bf e}_2,\quad [{\bf e}_2,{\bf e}_3]=-{\bf e}_1  \end{matrix} $ & $\frac
23 \delta^4_{~i}$
\\ \hline
\end{tabular}
\caption{The indecomposable four-dimensional algebras}
\label{table4dim}
\end{table}

Note that some of the parametric limits yields various other Lie
algebras. We have
\beq\begin{matrix}
\lim_{p\rightarrow 0}A^p_{4,2}&=&IV\oplus\mb{R},& &\lim_{q\rightarrow
0}A^{pq}_{4,5}&=& VI_h\oplus\mb{R}, \\
\lim_{p\rightarrow 0}A^{pq}_{4,6}&=&VII_h\oplus\mb{R},  & &
\lim_{q\rightarrow 0}A^q_{4,9}& = &A_{4,8}, \\ 
\lim_{q\rightarrow 0}A^q_{4,11}&= &A_{4,10}. & & & &
\end{matrix}
\eeq

Unfortunately there does not exist a simple expression, as  in
the three-dimensional case, for the structure
constants. This is perhaps the greatest obstacle to our
analysis. However, in several cases, the general form of the structure
constants can be worked out. 

Let $C^k_{~ij}$ be a special representation of a Lie algebra $\mc{A}$. We
define the space $\mc{W}(\mc{A})$ as
\beq
\mc{W}(\mc{A})=\left\{\widetilde{C}^k_{~ij}\big|
\widetilde{C}^k_{~ij}=\left({\sf A}^{-1}\right)^k_{~l}C^l_{~nm}{\sf
A}_{~i}^{n}{\sf A}_{~j}^{m},\quad {\sf A}\in GL(4,\mb{R})\right\}.
\eeq
Two elements in $\mc{W}(\mc{A})$ correspond to two isomorphic Lie
algebras. It would be convenient to have a specific
parametrisation of $\mc{W}(\mc{A})$ for all of the four-dimensional Lie
algebras. Unfortunately, such a parametrisation has not been found in the
four-dimensional case. The union of $\mc{W}(\mc{A})$ over all possible
Lie algebras in $n$-dimensions is called the \emph{variety of
$n$-dimensional Lie algebras}\footnote{For the case of complex Lie
algebras of low dimensions, see \cite{KN}.}. In dimensions two and three, the
structure of this variety is known, but in four dimensions and higher
it is not known in detail. 

However, for specific cases, $\mc{W}(\mc{A})$ can be found
explicitly. For instance, for the pure vector type algebra,
$A^{1,1}_{4,5}$ the structure constants are (see Table \ref{table4dim}) 
\beq
C^k_{~ij}=\delta^k_{~i}\delta^4_{~j}-\delta^k_{~j}\delta^4_{~i}.
\eeq 
Hence, $\mc{W}(A^{1,1}_{4,5})$ is the space of all non-zero vectors in
$\mb{R}^4$:
\beq
\mc{W}(A^{1,1}_{4,5})\cong \mb{R}^4\setminus \{0\}.
\eeq

\subsubsection{Non-unimodular Lie algebras}
Let us now investigate the algebraic way of looking at these Lie
algebras, following MacCallum \cite{MacC}. We will first investigate the ones that are
non-unimodular. These are exactly those who have non-zero trace:
$a_i\neq 0$. Let $\hat a_i$ be the unit vector
parallel to $a_i$. For the sake of simplicity, let us assume that our
frame is orientated so that $a_i=a\delta^4_{~i}$. Then we can
decompose the structure constants into 
\beq
C^A_{~B4} &= &\Theta^{A}_{~B}\nonumber \\
n^{ab} &=& \frac 12 C^a_{~ij}\epsilon^{ijbk}\hat{a}_k
\eeq
where $n^{ab}$ is a symmetric matrix.  The Jacobi identity implies
 $C^4_{~ab}=0$ and thus
\beq
n^{a4}=n^{4a}=0.
\eeq
Note that the matrix $n^{ab}$ only contributes to the trace-free part
of the structure constants. Using a ``sloppy'' notation, but practical
for our purposes, we can write the trace-free part as
\beq
D^A_{~BC} &=& \epsilon_{BCDk}n^{AD}\hat{a}^k \nonumber \\
D^A_{~B4} &=& \Theta^A_{~B}-a\delta^A_{~B}.
\eeq
We have still an $SO(3)$ orientation we can use to diagonalise
$n^{AB}$. Thus we can assume that
$n^{AB}=\text{diag}(n_1,n_2,n_3)$ by choosing a suitable frame. The
Jacobi identity reduces now
to
\beq
n_1(2D^1_{~14}-a)=n_2(2D^2_{~24}-a)=n_3(2D^3_{~34}-a)
&= &0 \nonumber \\
n_2\Theta^3_{~2}+n_3\Theta^2_{~3}=n_3\Theta^1_{~3}+n_1\Theta^3_{~1}=n_1\Theta^2_{~1}+n_2\Theta^1_{~2}
&=& 0.
\label{eqJacobiN}\eeq
Let us first assume that the rank of $n^{ab}$ is 3. This leads to
$D^A_{~A4}=3a/2$ which is a contradiction, since $D^a_{~bc}$ is
trace-free. Hence, we can assume that $n_3=0$. 

The classification now reduces to finding the eigenvalues $n_1$ and
$n_2$ and the matrix $\Theta^A_{~B}$. This is done in MacCallum
\cite{MacC};  we will take the simplest example in this case
(even though the dynamical behaviour of these types may be highly complex). Assume that the
rank of $n^{ab}$ is two. The further analysis splits in two cases, determined by the
sign of $n_1n_2$. Assume that $n_1,n_2 > 0$. This leads to
$D^1_{~14}=a/2$ and $D^2_{~24}=a/2$ (and hence $D^3_{~34}=-a$). Further
we get $\Theta^3_{~2}=\Theta^3_{~1}=0$ plus the constraint
$n_1\Theta^2_{~1}+n_2\Theta^1_{~2}=0$. This algebra is type
$A_{4,12}$, which MacCallum calls N22. 

The other Lie algebra with rank$(n^{ab})=2$ is found when
$n_1<0<n_2$. This corresponds to the decomposable algebra
$A_{2,1}\oplus A_{2,1}$. MacCallum calls this type N20. 

Hence, both of these algebras, $A_{4,12}$ and $A_{2,1}\oplus A_{2,1}$,
have (after choosing a orientation of frame) the following parameters:
\beq
n^{AB}=\text{diag}(n_1,n_2,0), \quad D^{A}_{~B4}=\begin{bmatrix} 
\frac a2 & -\frac{n_1}{n_2}\Theta^2_{~1} & \Theta^1_{~3} \\
\Theta^2_{~1} & \frac a2 & \Theta^2_{~3} \\
0 & 0 & -a \end{bmatrix}.
\eeq 
Doing this analysis for all of the remaining cases (for rank 1 and 0)
we can find the parameter space for these models. In the appendix, the
general form of the matrix $\Theta^{A}_{~B}$ is listed for all of the
non-unimodular Lie algebras (using a particular choice of gauge). For example, in the
case where $n_1=n_2=0$, eq. (\ref{eqJacobiN}) vanishes
identically. Thus it remains to classify the different invariant
properties of the matrix $\Theta^A_{~B}$. These invariant properties
determine the Lie algebra type. 

\subsubsection{Unimodular algebras}
The unimodular algebras are defined by vanishing trace: $a_i=0$. For
these algebras the following theorem holds:
\paragraph{Theorem (Farnsworth and Kerr)} {\sl For a four-dimensional
unimodular Lie algebra there will either exist a $p_a$ such that 
\beq
C^a_{~bc}=\Theta^a_{~[b}p_{c]}, \quad \Theta^a_{~b}p_{a}=0,
\label{U1}\eeq
or there exists no such $p_a$, and there exists a non-zero $\ell^c$ such
that }
\beq
C^a_{~bc}\ell^c=0.
\label{U3}\eeq
A simple and geometric proof of this is given in MacCallum \cite{MacC}. 
MacCallum calls the algebras obeying (\ref{U1}) and (\ref{U3}), U1 and
U3 respectively. 

In the class U1, we can choose an orientation such that
$p_a=p\delta_{~a}^4$. Now the remaining part to be classified is the
matrix $\Theta^{A}_{~B}$. Since the structure constants are
trace-free, it follows that $\Theta^A_{~A}=0$. 

The class U3 splits in two; U3I and U3S. Each of these subclasses has
two members each. The members of the class U3I, $A_{4,8}$ and
$A_{4,10}$, can both be seen as parametric limits of a
non-unimodular model. The members of class U3S are semi-simple and are
the decomposable algebras VIII$\oplus \mb{R}$ and IX$\oplus \mb{R}$. 
 
\subsection{Examples and solutions}
\label{sect:ex}
\subsubsection{The types $A_3\oplus\mb{R}$}
It is interesting to investigate the trivial expansion of the Bianchi
types explicitly. These models correspond to the simplest
Kaluza-Klein models. They have a single extra dimension, and since
this dimension is homogeneous, it can be compactified into a
circle. 

We need to find all the structure constants
possible for these models. There are  two equivalent ways
of doing this. Either we can do it purely algebraically, or we can do
it from a geometrical point of view. We will choose the latter.

In principle, the extra dimension can be
tilted. Let us choose the spatial vierbein to have three vectors
spanning the vectorspace $A_3$. Hence, we let three vectors have 
a Bianchi-type algebra, while the fourth vector is orthogonal to
$A_3$. Let these three vectors which span $A_3$ be denoted ${\bf
e}_B$, $B=1,2,3$. We can now write
\beq
C^D_{~AB}=\epsilon_{ABC}n^{CD}+\delta^D_{\phantom{D}A}\tilde{a}_B-\delta^D_{\phantom{D}B}\tilde{a}_A
\eeq
with the requirement
\beq
\tilde{a}_Dn^{CD}=0.
\eeq
We know there exists a vector ${\bf u}$ which commutes with ${\bf
e}_B$. This vector is linearly independent of ${\bf e}_B$ so we can
write ${\bf e}_4=\lambda^A{\bf e}_A+\lambda^4{\bf u}$ where $\lambda^4\neq 0$. Hence we get
\beq
[{\bf e}_A,{\bf e}_4]\equiv C^D_{~A4}{\bf e}_D=\lambda^BC^D_{~AB}{\bf
e}_D
\eeq
Thus the structure constants can be written
\beq
C^D_{~A4}=\lambda^BC^D_{~AB}, \quad C^4_{~ab}=0.
\eeq
The trace of the structure constants can now be calculated
\beq
a_B&=& \frac 13 C^i_{~iB}=\frac 23 \tilde{a}_B \nonumber \\
a_4&=& \frac 13 C^i_{~i4}=\frac 23\lambda^D\tilde{a}_D=\lambda^Da_D
\eeq
and the trace-free part is
\beq
D^A_{~BC}&=&\epsilon_{BCD}n^{DA}+\frac
12\left(\delta^A_{\phantom{S}B}a_C-\delta^A_{\phantom{A}C}a_B\right) \nonumber \\
D^A_{~B4}&=&\lambda^C\epsilon_{BCD}n^{DA}-\frac
32\lambda^Aa_B+\frac 12\delta^A_{~B}a_4 \nonumber \\
D^4_{~A4} &=& a_A.
\eeq
We still have an $SO(3)$ choice of gauge to fix the orientation of the
vectors ${\bf e}_B$, hence, we can choose a orientation where the
matrix $n^{CD}$ is diagonal. 

\paragraph{Class A models:} 
Let us turn our attention to the Class A models. The class A models are
characterised by 
\[ a_B=0. \]

The Jacobi equations and the $(0,i)$ field equations lead to a set of
constraints which must be satisfied. Let us choose a frame where $n^{AB}$ is
diagonal. The constraint equations can now be satisfied with (or a
frame can be chosen)
\beq
\sigma_{AB}= &0 &=\Omega_{AB} \quad (A\neq B) \nonumber \\
\lambda_A = &0& \nonumber \\
\sigma_{4A} &=& \Omega_{4A}.
\eeq

Note that we have not assumed $\sigma_{4A}=0$, and in general we have
not enough gauge freedom to put the shear tensor into diagonal
form. The case I$\oplus\mb{R}$ is the only case where we can diagonalise the
shear vector completely. In the other models there will still remain
some off-diagonal shear components. 
In the absence of anisotropic stress, we will get further constraints
from the off-diagonal shear equations. These can be satisfied with
\beq
\Omega_{4A}=0=\sigma_{4A}.
\eeq
Thus, in this case the shear must be diagonal. Henceforth we will
assume that this is the case. 

The tensor $b_{ab}$ reduces to the  form
\beq
b_{AB} &=
&2n_{AC}n^{C}_{\phantom{C}B}-n^{C}_{{\phantom{C}C}}n_{AB}+\frac
12\delta_{AB}\left[\left(n^{C}_{{\phantom{C}C}}\right)^2-2n_{AC}n^{C}_{\phantom{C}B}\right] \nonumber \\
b_{4a} &=&0
\eeq
The independent shear equations are now
($\sigma^4_{~4}=-\sigma^A_{~A}$)
\beq
\dot{\sigma}_{AB}+\theta\sigma_{AB}+b_{AB}-\frac
14\delta_{AB}b^C_{~C}= 0
\eeq
while the remaining Jacobi equations are 
\beq
\dot{n}^{AB}+\frac 14\theta
n^{AB}+\sigma^C_{\phantom{C}C}n^{AB}-2\sigma^{(A}_{\phantom{(A}C}n^{B)C}=0.
\eeq
Note that these are only valid in the choice of gauge mentioned
above. 

\paragraph{Example: A II$\oplus\mb{R}$ perfect fluid solution.} 
Let us consider a specific case. Assume that the matrix $n^{AB}$ has
only one non-zero eigenvector. We assume that 
\beq
n^{AB}=\text{diag}(n,0,0)
\eeq
which corresponds to the II$\oplus\mb{R}$ Lie algebra.
We introduce new variables $\sigma_{1,2,3}$ by
\beq
\sigma_{11} &=& 3\sigma_1 \nonumber \\
\sigma_{22} &=& -\sigma_1-\sqrt 2\sigma_2+\sqrt 6\sigma_3  \nonumber
\\
\sigma_{33} &=& -\sigma_1-\sqrt 2\sigma_2-\sqrt 6\sigma_3  \nonumber
\\
\sigma_{44} &=& -\sigma_1+2\sqrt 2\sigma_2.
\label{eqsigmas}\eeq
The equations of motion now reduce to 
\beq
\dot{\sigma}_1+\theta\sigma_1 +\frac{5}{24}n^2 &=& 0 \nonumber \\
\dot{\sigma}_2+\theta\sigma_2 +\frac{1}{6\sqrt{6}}n^2 &=& 0 \nonumber
\\
\dot{\sigma}_3+\theta\sigma_3  &=& 0 \nonumber \\
\dot{n}+\frac 14\theta n-5\sigma_1 n-2\sqrt{2}\sigma_2n &=&0.
\eeq
In addition to these,  Raychaudhuri's equation and the energy
conservation equation must be fulfilled. We will assume in this
example 
that the perfect fluid obeys a $\gamma$-law equation of state:
$p=(\gamma-1)\rho$. 

There is a specific case where we can solve the equations of motion exactly. We
can search for a self-similar solution with the properties that
$\sigma_{1,2,3}\propto t^{-1}$, $\rho\propto t^{-2}$ and $\theta
\propto t^{-1}$. By doing this, we can find a solution which
is the II$\oplus \mb{R}$ version of the Collins-Stewart type II perfect
fluid solution. This solution is given by
\beq
\sigma_1 &=& -\frac{5}{66\gamma}(2\gamma -1)t^{-1} \nonumber \\
\sigma_2 &=& -\frac{\sqrt{2}}{33\gamma}(2\gamma -1)t^{-1} \nonumber \\
\sigma_3 &=& 0 \nonumber \\
n^2 &=& \frac{4}{11\gamma^2}(2-\gamma)(2\gamma-1)t^{-2} \nonumber \\
\rho &=& \frac{3}{11\gamma^2}(6-\gamma)t^{-2} \nonumber \\
\theta &=& \frac{2}{\gamma}t^{-1}
\eeq
where $1/2\leq \gamma \leq 2$.\footnote{It should be noted that in
4+1 dimensions, radiation has $\gamma=5/4$, and the strong energy
condition requires $\gamma>1/2$. Hence, inflation can occur for $\gamma<1/2.$} The metric can written as
\beq
ds^2&= &-dt^2+t^{\frac{2}{11\gamma}(8-5\gamma)}\left(dx+\frac{2c}{\gamma}ydz\right)^2
\nonumber \\ &&
+t^{\frac{2}{11\gamma}(3\gamma+4)}\left(dy^2+dz^2\right)+t^{\frac{2}{11\gamma}(6-\gamma)}dw^2
\eeq
where $c^2=(2-\gamma)(2\gamma-1)/11$. Note that if $\gamma> 8/5$ -- if
the matter is stiff enough -- one of the dimensions will contract
to an arbitrary small size.

\subsubsection{Solutions to the model $A^{1,1}_{4,5}$}
This model is of pure vector type and is particularly easy. We have
already worked out the space of structure constants for this
model. The structure constants can be characterised by a non-zero
vector $a_i$ in $\mb{R}^4$. We choose an orientation of the frame so that
${\bf e}_1$ is aligned with $a_i$. Hence, $a_i=\delta_{~i}^1a$ for
$a\neq 0$.  

The Jacobi equations (\ref{Jacobi1}) and (\ref{Jacobi2}) are satisfied
by construction, and eq. (\ref{Jacobi4}) is trivially satisfied. Equation
(\ref{Jacobi3}) leads to 
\beq
\dot{a}+\frac 14\theta a+\sigma_{11}a=0
\eeq
and
\beq
\Omega_{12}=\sigma_{12},\quad \Omega_{13}=\sigma_{13}\ \ \text{and
}\Omega_{14}=\sigma_{14}
\eeq
Eq. (\ref{R0a}) equation tells us that 
\beq
\sigma_{a1}a=0\Rightarrow \sigma_{a1}=0.
\eeq
Hence, the shear has only non-zero components in the 2, 3 and 4
directions. We still have an unused freedom of choosing the
orientation of the vectors ${\bf e}_i$ for $i=2,3$ and 4. Hence, we
can choose a frame where the shear $\sigma_{ab}$ is diagonal. 

Henceforth, we will also assume  $\Lambda=0=\pi_{ab}$ and that
the fluid obeys the barotropic equation of state $p=(\gamma-1)\rho$. 

The dynamical systems approach has proved to be a powerful tool in
cosmology (see for example \cite{DynSys}). Let us use this method to solve the complete system of
equations in the $A^{1,1}_{4,5}$ model. We parametrise the shear with
\beq
\sigma_{ab}=\text{diag}(0,-2\sigma_+,\sigma_++\sqrt{3}\sigma_-,\sigma_+-\sqrt{3}\sigma_-)
\eeq
and introduce a new time coordinate with
\beq
\frac{dt}{d\tau}=\frac{4}{\theta}.
\eeq
Introducing the expansion normalised variables
\beq
\Sigma_{\pm}=\frac{2\sqrt{2}\sigma_{\pm}}{\theta},\quad
\mc{A}=\frac{4a}{\theta},\quad \Omega=\frac{8\rho}{3\theta^2}
\eeq
the equations of motion can be written as
\beq
\Sigma'_{\pm} &=& (q-3)\Sigma_{\pm} \nonumber \\
\mc{A}' &=& q\mc{A} \nonumber \\
\Omega' &= &2[q-(2\gamma-1)]\Omega
\eeq
where 
\beq
q &=& 3\Sigma+(2\gamma-1)\Omega \nonumber \\
\Sigma&=& \Sigma^2_++\Sigma^2_-\nonumber \\
1 &=& \Sigma+\mc{A}^2+\Omega.
\eeq
The last of these is the constraint equation, and $q$ is the deceleration parameter defined by
$\theta'=-(1+q)\theta$. 

The system of equations turns out to be quite simple, which
makes it possible to solve these equations exactly. The solutions are 
\beq
\Sigma_{\pm} &=&
\frac{p_{\pm}e^{-3\tau}}{\left(A^2e^{-6\tau}+e^{-2(2\gamma-1)\tau}+K^2\right)^{1/2}}
\nonumber \\
\Omega &=&
\frac{e^{-2(2\gamma-1)\tau}}{A^2e^{-6\tau}+e^{-2(2\gamma-1)\tau}+K^2}
\nonumber \\
\mc{A} &=&
\frac{K}{\left(A^2e^{-6\tau}+e^{-2(2\gamma-1)\tau}+K^2\right)^{1/2}}
\eeq
where $p_+^2+p_-^2=A^2$. Note that the solutions are Kasner-like near the initial
singularity. At late times, the universe approaches the 5D Milne
universe if $1/2<\gamma< 2$ and the flat 5D FRW if $0\leq \gamma<
1/2$. 

We introduce the functions
\beq
N(t) &\equiv &\left(A^2+t^{2-\gamma}+K^2t^{3/2}\right)^{\frac
12} \nonumber \\ {}
\beta(t)&\equiv &
\int\frac{A dt}{t\cdot N(t)}.
\eeq
The metric for the solutions can now be written
\beq 
ds^2&=&-\frac{dt^2}{N(t)^2}+t^{1/2}\bigg\{dx^2+e^{-\frac{K}{2}x}\bigg[e^{-\sqrt{2}\cos\left(\phi\right)\beta(t)}dy^2\nonumber\\
&& +e^{\sqrt{2}\cos\left(\phi+\pi/3\right)\beta(t)}dz^2+e^{\sqrt{2}\cos\left(\phi-\pi/3\right)\beta(t)}dw^2\bigg]\bigg\}.
\eeq

\subsubsection{The four-dimensional ${\sf Nil}^4$ case}
The only indecomposable Lie algebra in four dimensions that is nilpotent, is the type
$A_{4,1}$. This algebra is the Lie algebra of the four-dimensional
nil-geometry, denoted by ${\sf Nil}^4$. For the sake of illustration we
will determine the possible structure constants for this model. 

What we would like to have is a representation of the space
$\mc{W}(A_{4,1})$. But finding the whole space is by no means
necessary. We have a gauge freedom which corresponds to an
$O(4)$-rotation of the spatial frame. By the Gram-Schmidt process, we can show that every ${\sf A}\in GL(n,\mb{R})$ can
be written as
\beq
{\sf A}={\sf P}\cdot{\sf R}
\eeq
where ${\sf R}\in O(n)$ and ${\sf P}\in PT(n)$ where $PT(n)\subset GL(n,\mb{R})$ is the
group of all upper-triangular matrices
with positive entries along the diagonal. Hence, by choosing a
suitable frame, we can ``gauge away'' the rotation matrix ${\sf
R}$. It suffices therefore to look at the space 
\beq
\mc{P}(\mc{A})=\mc{W}(\mc{A})/ O(4)= \left\{\widetilde{C}^k_{~ij}\big|
\widetilde{C}^k_{~ij}=\left({\sf P}^{-1}\right)^k_{~l}C^l_{~nm}{\sf
P}_{~i}^{n}{\sf P}_{~j}^{m},\quad {\sf P}\in PT(4)\right\}.
\eeq
$PT(4)$ is a Lie group, thus if ${\sf P}\in PT(4)$ then ${\sf
P}^{-1}\in PT(4)$. From Table \ref{table4dim} we have
\beq
C^1_{~24}=1,\quad C^2_{~34}=1.
\eeq
By calculating $\mc{P}(\mc{A})$ we find that there are essentially only three non-zero
commutators. These are
\beq
C^1_{~24},\quad C^1_{~34} \quad \text{and}\quad C^2_{~34}.
\eeq
(see also the appendix where the $A_{4,1}$ can be seen as the
$a\longrightarrow 0$ limit of $A_{4,4}$.) 
The Lie algebra is trace-free, so $C^k_{~ij}=D^k_{~ij}$.
Note that in the general case, the three-dimensional Ricci tensor is
not diagonal. Hence, if the shear is to be diagonal, then we have to
set either $D^1_{~24}$ or $D^2_{~34}$ to zero if no anisotropic stress
is present. Let us assume therefore that $\pi_{ab}=0$, $C^1_{~34}=0$,
$\Omega_{ab}=0$ and $\sigma_{ab}$ diagonal. 

We introduce the two curvature variables 
\beq
N_1=D^1_{~24},\quad N_2=D^2_{~34}.
\eeq
So in this model, there will be 5 variables left to describe the
geometry of the spatial hypersurfaces. These
are
\beq
N_1,\quad N_2,\quad \sigma_{1,2,3}
\eeq
where $\sigma_{1,2,3}$ are defined in eq. (\ref{eqsigmas}). Together
with the matter equations, the evolution equation for these variables
can now be written down. 

In the case of a $\gamma$-law perfect fluid, one can also find
a specific self-similar solution. The metric for this ${\sf Nil}^4$
solution is
\beq
ds^2 &=& -dt^2+t^{\frac
25(4-3\gamma)}\left(dx+\frac{2c}{\gamma}ydw\right)^2+t^{\frac
25(3-\gamma)}\left(dy+\frac{2c}{\gamma}zdw\right)^2 \nonumber \\
&&+t^{\frac
25(\gamma+2)}dz^2+t^{\frac
25(3\gamma+1)}dw^2
\eeq
where $c^2=(2-\gamma)(2\gamma-1)/10$ and $1/2\leq \gamma \leq 2$. The
matter-density of the perfect fluid is
\beq
\rho=\frac{3(3-\gamma)}{5\gamma^2t^2}.
\eeq
For this solution, if $\gamma\geq 4/3$ then there will be one
contracting direction. 
However, this
solution is only a particular solution, corresponding to a fixed point
in the dynamical system. The more general behaviour and the nature of
this solution for the ${\sf Nil^4}$-world will be the subject of a future
work. 

\subsection{Exceptional models $A^*_{4}$}
For the Bianchi models in 3+1 dimensions, there is one exceptional
case VI$^*_{-1/9}$ for which one of the $R_{0a}$-equations vanish
identically; it can have one extra shear degree of freedom. 

In the 4+1 dimensional case,  we have many such exceptional cases. By
inspecting eq. (\ref{R0a}), we see that we can have an additional
shear degree of freedom in the following cases:
\beq
A^p_{4,2}: && p=-1 \nonumber \\
A^{pq}_{4,5}: && q=-\frac{1+p}{2}\quad (\text{or with $q,p$
interchanged})\nonumber \\
A^{pq}_{4,6}: && p=-q. \nonumber 
\eeq
We will denote these models which have this additional shear degree of
freedom with $A^*_4$. Note that the case VI$^*_{-1/9}\oplus \mb{R}$ is
obtained in the limit of $A^{pq*}_{4,5}$ as $p\rightarrow 0$:
\beq
\lim_{p\rightarrow 0}A^{pq*}_{4,5}= VI^*_{-1/9}\oplus \mb{R}.
\eeq
The other exceptional models are special for the 4+1 dimensional case
and have no 
3+1 dimensional analogue. Note also that in the cases
$A^{-\frac 13,-\frac 13*}_{4,5}$ and $A^{1,-1*}_{4,5}$ we can have \emph{two} additional shear degrees
of freedom. However, the total number of parameters remains the same
because in these cases we have one less commutator function (see
appendix)\footnote{This is because we have an additional symmetry in
these cases which makes it possible to gauge away one more 
commutator function.}. 

Hence, for a given Lie algebra, these exceptional models have more
degrees of freedom than any of the other models.

\section{Compactification}

As we now have presented all the homogeneous models, we note that
most of the models are non-compact. The Kaluza-Klein mechanism needs a
small and compact dimension to work, and hence, all of the models
which do not have a compact dimension cannot be a proper model for the
Kaluza-Klein mechanism. However, in many of the models we can construct
compact versions of the models. In a Kaluza-Klein model one usually
have a product space
\beq
M\times N
\eeq
where $M$ is the 3+1 dimensional base space and $N$ is a compact
space. For example, in string theory $N$ is a Calabi-Yau manifold with
3 complex dimensions. In our model, which has in total 5 dimensions,
$N$ must be one-dimensional, and thus $N=S^1$.  

However, one could instead imagine a more general model where the extra
dimension is small, compact and perhaps ``twisted''. Hence, we do not
necessarily need to constrain ourself to a product space, but we can allow for
a more general space. We can assume that our space is a
\emph{fiberbundle} $P$ with compact fibers. For a fiberbundle there will also exist a
projection map $\pi: P\mapsto M$. $M$ is called the base space, and in
our context this is our four-dimensional spacetime. We demand that for a $p\in P$ we
have 
\beq
\pi^{-1}(p)=S^1.
\eeq
 Locally we will not see the difference between these
``twisted'' spaces and the product spaces; for any $p\in P$ there will exist an open neighbourhood
$U\in M$ such that 
\beq
U\times S^1 \subset P.
\eeq
The difference will be at a global scale. In a cosmological setting,
the space $M$ will be of cosmological size,
and hence, we have to go to cosmological scales to see the difference
between the twisted and the un-twisted versions. Notwithstanding, in the very early
universe the cosmological scales may have been comparable to the small
extra dimensions. Thus these twisted spaces may have had an important
impact on the early evolution of the universe. 

To construct a compact dimension we can proceed as follows. We find a
discrete subgroup 
$\Gamma \subset \text{Isom}(P)$ which acts freely and properly
discontinuously on $P$. We identify now points $p$ and $q$ for which
there exist a $\gamma\in\Gamma$  such that $p=\gamma(q)$. There are usually
many such groups for a homogeneous manifold, but not always. Also in
many cases, the compactification radius can vary as we move along the
base space.  
For the simply transitive models this compactification can always be
done. These models have four linearly independent Killing vectors, ${\mbold\xi}_i$,
which act freely on the spatially homogeneous hypersurfaces. Hence,
for any Killing vector ${\mbold\xi}$ which is a linear combination of the
${\mbold\xi}_i$'s we can -- through exponentiation
$\phi=\exp{\mbold\xi}$ -- find such a freely and properly
discontinuous acting group
$\Gamma$ generated by the element $\phi$.

In the decomposable models, the compactification depends on the single
extra dimension, which is trivial. In many of these cases, we can also 
compactify the base space as well, and in three dimensions this has
some very interesting consequences
\cite{as,fik,Kodama1,BK1,BK2,Kodama2}. In particular, due to Mostow's
rigidity theorem, all compact hyperbolic spaces of dimension three or
higher, are rigid; i.e. they do not allow an anisotropic
expansion. For the Bianchi types III and VIII (and hence types
III$\oplus\mb{R}$ and VIII$\oplus\mb{R}$ as well) the effect of
compactification is quite the opposite; they can have a unbounded
number of free parameters. The compactification induces so-called
moduli parameters which will increase the number of free
parameters. For example, the vacuum type I$\oplus \mb{R}$ model has 2
free parameters, while the compactified version of type I$\oplus\mb{R}$ into a four-torus $T^4$ has 2+16 free
parameters\footnote{For the 3+1 dimensional case this is illustrated
in \cite{sig1}.} . This  may have a significant effect on a quantum theory of
gravity and on quantum
cosmological models.

We would leave the compactification question open, but we will provide
with an example in which we compactify both the base space and its
fibers. 

\paragraph{Compactification of the nilpotent geometries.}
We start by considering a three-dimensional torus $T^3$. The torus can
be constructed from the usual $\mb{E}^3$ by identification under a
discrete group of translations. 

Consider the product between the torus and the finite interval $[0,1]$
\beq
T^3\times [0,1].
\label{T3times01}\eeq
The torus can be viewed upon as the  unit cube in $\mb{E}^3$ with the usual
identification of the  boundary. We will now identify the two tori on boundary
of (\ref{T3times01}) as follows. Take a matrix ${\sf A}\in
SL(3,\mb{Z})$ (the special linear group with integer
entries). $SL(3,\mb{Z})$ is the mapping class group of $T^3$, hence,
the mapping ${\sf A}(T^3)$ maps the torus isometrically onto
itself. To obtain a compact space we can therefore identify the
boundary $T^3\times\{0\}$ with $T^3\times\{1\}$ under the mapping of
${\sf A}$. Thus if $p, q \in T^3$
then
\beq
\{p\}\times\{0\} \sim \{q\}\times\{1\}, \text{ iff } {\sf A}(q)=p.
\eeq 
We have now constructed a compact manifold $M\cong T^3\times [0,1] /
\sim$.  If the matrix ${\sf A}$ is the identity matrix, then $M\cong
T^4$ and hence, is of type $I\oplus\mb{R}$. The three other
possibilities are the two  nilpotent spaces\footnote{${\sf Nil}^3\times
S^1$ has the type $II\oplus \mb{R}$  while the ${\sf
Sol}^3\times S^1$ has a type $VI_0\oplus \mb{R}$ Lie algebra.}  ${\sf Nil}^3\times S^1$
and ${\sf Nil}^4$, and the solvable case ${\sf Sol}^3\times S^1$. 

We define the characteristic polynomial $p_{\sf
A}(\lambda)$ by
\beq
p_{\sf
A}(\lambda)\equiv\det\left({\sf A}-\lambda {\sf 1}\right).
\eeq
The roots of $p_{\sf
A}(\lambda)$ determine whether we have a nilpotent group or not. 

We have a nilpotent geometry if all three roots of ${\sf A}$ are equal to 1. If
this is not the case, then we have the solvable case ${\sf Sol}^3\times S^1$.
All of the nilpotent geometries in four dimensions
can be compactified completely in this way, depending on the
reducibility of the matrix ${\sf A}$. 

\section{Conclusion and Outlook}
In this paper we have classified all spatially homogeneous
cosmological models of dimension 4+1 where the spatial hypersurfaces are connected and
simply connected. We found five multiply transitive models which cannot
 be seen upon as special cases of the simply transitive
models. These are in some sense the four-dimensional versions of the
Kantowski-Sachs (KS) model. Three of these models were directly linked
to the KS case since they consisted of products of manifolds where at
least one of the components was a sphere $S^2$. 

For the simply transitive models, the classification of the four
dimensional Lie algebras provides us with all the possible
models. Among these models we should expect many interesting
phenomena. For example, they may give us an understanding of how
extra dimensions can affect the evolution of our four dimensional
universe. The idea of extra dimensions is by no means a new idea, but
higher-dimensional cosmologies seems to be little understood. 

A special class of solutions to the field equations are
particularly interesting. The plane-wave solutions are known to be solutions
describing gravitational waves propagating through spacetime. A feature of
these solutions is that they possess an extra symmetry in addition to
those arising from 
the requirement of spatial homogeneity. These solutions have a null Killing
vector. The total symmetry groups are therefore higher than the generic
solution of the homogeneous field equations. These solutions also have
a particular interest in string theory because they admit
supersymmetry. In 4+1 dimensions there exists a large class of plane
wave solutions. As a matter of fact, there is a five parameter family
of vacuum plane-wave solution in 4+1 dimensions. Some of them
generalises the known plane wave 
solutions in 3+1 dimensions, others are new and special for 4+1
dimensions. These solutions will be the subject of a future work.

A recent work which should be mentioned in relation to this, is
a work by De Smet \cite{DeSmet}. In this work, a Petrov
classification of algebraically special five-dimensional spacetimes
is constructed. However, 
what special role these spacetimes may have for  multidimensional
cosmology is not known.

\section*{Acknowledgments}
The author is deeply grateful to J.D. Barrow for useful comments
and suggestions on this work. 

This work was founded by the Research Council of Norway.

\section*{Appendix: The structure constants for the indecomposable Lie
algebras}

We will here give a list of all the structure constants for the
indecomposable ones. We have chosen a specific gauge (orientation of frame) to simplify the
expressions and clarify their structure. We write them in terms of the
two 
matrices $n^{AB}=\text{diag}(n_1,n_2,0)$ and
$C^{A}_{\phantom{A}B4}=\Theta^A_{\phantom{A}B}$. Note also that the
obtained bound for the structure constants may in some cases not
hold globally.

\subsubsection*{rank$(n^{AB})=2:\ n_1,n_2\neq 0$}
\beq 
\Theta^{A}_{\phantom{A}B}=\begin{bmatrix} 
\frac 32a & -\frac{n_1}{n_2}\Theta^2_{~1} & \Theta^1_{~3} \\
\Theta^2_{~1} & \frac 32a & \Theta^2_{~3} \\
0 & 0 & 0 \end{bmatrix}.
\eeq 
\begin{itemize}
\item{$A_{4,12}$: } $0<n_1,n_2$
\item{$A_{2,1}\oplus A_{2,1}$:} $n_1<0<n_2$.
\end{itemize}
\subsubsection*{rank$(n^{AB})=1:\ n_1\neq 0,\ n_2=0$}

\beq 
\Theta^{A}_{\phantom{A}B}=\begin{bmatrix} 
\frac 32a & \Theta^1_{~2} & \Theta^1_{~3} \\
0 & \Theta^2_{~2} & \Theta^2_{~3} \\
0 & \Theta^3_{~2} & \Theta^3_{~3} \end{bmatrix}.
\eeq 
\begin{itemize}
\item{$A^q_{4,9}$:} $\Theta^3_{~2}=0$,\  $\Theta^2_{~2}>0$,\ 
$q\Theta^2_{~2}=\Theta^3_{~3}$,\ 
$\Theta^2_{~2}+\Theta^3_{~3}=\frac{3}{2}a$\\
$q=1\Rightarrow \Theta^2_{~3}=0$.
\item{$A_{4,7}$:} $\Theta^3_{~2}=0$, $\Theta^2_{~3}>0$,
$\Theta^2_{~2}=\Theta^3_{~3}>0$
\item{$A_{4,11}^q$:} $\text{Tr}({\sf T})=\frac 32 a$,\ $4q^2
\text{det}({\sf T})=\left(\text{Tr}({\sf T})\right)^2(1+q^2)$ where 
\beq
{\sf T}=\begin{bmatrix} \Theta^2_{~2} & \Theta^2_{~3} \\
 \Theta^3_{~2} & \Theta^3_{~3} \end{bmatrix}
\label{eqTheta23}\eeq
\end{itemize}
\subsubsection*{rank$(n^{AB})=0:\ n_1=n_2=0$}

\beq 
\Theta^{A}_{\phantom{A}B}=\begin{bmatrix} 
\Theta^1_{~1} & \Theta^1_{~2} & \Theta^1_{~3} \\
0 & \Theta^2_{~2} & \Theta^2_{~3} \\
0 &  0  & \Theta^3_{~3} \end{bmatrix}, \quad \Theta^A_{~A}=3a.
\eeq 
\begin{itemize}
\item{$A_{4,5}^{pq}$:} $\Theta^1_{~1}>0$,\
$p\Theta^1_{~1}=\Theta^2_{~2}$, \ $q\Theta^1_{~1}=\Theta^3_{~3}$.\\
$p=1\Rightarrow \Theta^1_{~2}=0$, $\quad p=q\Rightarrow
\Theta^2_{~3}=0$, $\quad p=q=1\Rightarrow \Theta^1_{~3}=0$
\item{$A_{4,2}^{p}$:} $\Theta^2_{~2}=\Theta^3_{~3}\neq 0$, \
$\Theta^1_{~1}=p\Theta^2_{~2}$,\ $\Theta^2_{~3}>0$, \\
$p=1\Rightarrow \Theta^1_{~2}=0$.
\item{$A_{4,3}$:} $\Theta^2_{~2}=\Theta^3_{~3}=0$,\ $\Theta^2_{~3}>0$
\item{$A_{4,4}$:} $\Theta^1_{~1}=\Theta^2_{~2}=\Theta^3_{~3}=a$,\ $\Theta^2_{~3}>0$
\item{$A_{4,6}^{pq}$:} $\Theta^1_{~1}=3ap/(p+2q)\neq 0$, $\text{Tr}({\sf
T})=6aq/(p+2q)>0$, \ $4q^2
\text{det}({\sf T})=\left(\text{Tr}({\sf T})\right)^2(1+q^2)$ where
${\sf T}$ is defined in eq. (\ref{eqTheta23}).
\end{itemize}
The rest of the non-unimodular Lie algebras can be obtained by taking
various parametric limits of the above algebras. 

Letting the trace go to zero, leads to
\beq
a &\longrightarrow & 0 \nonumber \\
A_{4,4}&\longrightarrow& A_{4,1} 
\eeq
and hence, all the indecomposable Lie algebras and the non-unimodular
algebras can be extracted from the above. The remaining ones, the
decomposable of class A, are included in the analysis in section \ref{sect:ex}.

\end{document}